\documentstyle[12pt,psfig]{article}
\begin{document}
\date{}
\bibliographystyle{unsrt}
\title{Design of Copolymeric Materials}
\author{Tanya Kurosky and J.M.Deutsch\\
University of California, Santa Cruz, U.S.A.}
\maketitle
\abstract{ We devise a method for designing materials
that will have some desired structural characteristics.
We apply it to multiblock copolymers that have two
different types of monomers, A and B.  We show how
to determine what sequence of A's and B's should be
synthesised in order to give a particular structure
and morphology.  
Using this method in conjunction with the theory of 
microphase separation developed by Leibler, we  
show it is possible to efficiently search for a
desired morphology.
The  method is quite general and can be extended to design
isolated heteropolymers, such as proteins, with desired
structural characteristics. We show that by making
certain approximations to the exact algorithm, a method recently 
proposed by Shakhnovich and Gutin is obtained. The problems with this method
are discussed and we propose an improved approximate algorithm
that is computationally efficient.}

\newpage
The problem addressed in this letter is the following.
Is there an efficient method for designing a material
with a particular morphology or structure? We develop
a systematic approach to this problem that we
illustrate for the design of copolymeric materials.

Structures of single chains in solution have also
been extensively studied, often in relation to the
important biological question of how to determine
the structure of a protein from its
sequence~\cite{creighton,dill}. Work on the design
of a chemical sequence which has a desired three
dimensional structure has also been recently
considered~\cite{chan}. For the two dimensional
model of Dill et al.~\cite{dill} it has been possible
to devise a set of rules that achieve a desired
tertiary structure~\cite{yue}.  In three dimensions,
much less is known. Ad hoc methods have been 
attempted\cite{shakhnovich1,shakhnovich2}
but recent tests have shown that they are not entirely 
efficacious\cite{whoops}.

To illustrate our general method, we will apply it
to design of block copolymers, which are 
polymers made out of more than one chemical species.  
We will consider
copolymeric systems made up of two constituent 
types of monomers denoted A and B.
The phase diagram of such
materials has been studied as a function of the Flory
interaction parameter $\chi$ and the lengths of the
segments of A and B. Lamellar, hexagonally closed packed,
body centred cubic\cite{leibler},
and gyroid\cite{matsen} phases have been predicted.
Experimentally copolymers have been found to exhibit
a variety of different structures, sometimes referred
to as microphases. 

Suppose we would like to design a new phase that
has a given symmetry. Until now, it was necessary
to do an inefficient search through the phase diagram
of the system in order to find the desired symmetry.
We show below that we can find a function that when
minimised can home in on the correct structure. We
then implement this practically
in the framework of the theory of copolymers developed
by Leibler\cite{leibler}. 
We then turn to the 
problem of protein design and discuss how the method
we developed can be used in this context.

First we wish to determine the correct function to minimise in order
to obtain the best sequence corresponding to the desired morphology.
We begin with a formulation of the general problem that we wish to solve.
Consider a system with coordinates denoted by $\Gamma$ and with a
chemical sequence denoted by $S$.  
We can
define a function which tells us whether each structure $\Gamma$
is in the
desired set of structures. Call this $P_{struct}(\Gamma )$.
It is a constant
if $\Gamma$ belongs to the class of desired structures and $0$
otherwise. In practice we will express $P_{struct}(\Gamma )$ in terms
of the clamping potential defined below.
Consider next the probability that a sequence $S$ gives
a desired structure. Since this is a conditional probability
we denote it by $P(S|struct)$. We wish to find the maximum of
$P(S|struct)$ over all sequences $S$  keeping the
structure fixed. To uniquely define $P(S|struct)$, we have to
choose the {\it a priori} probability of choosing an arbitrary
sequence $S$. The simplest choice is that it is uniform, that is
$P(S)$ is constant. 
We consider the system in equilibrium at a finite temperature
so the probability that a sequence $S$ has structure $\Gamma$ is
given by the Boltzmann factor  
$P(\Gamma | S) = \exp {(-[H_S(\Gamma)-F_o(S)]/T)}$, 
where  $H_S(\Gamma)$ is the
Hamiltonian of the system with a particular sequence $S$,
and $F_o(S)$ is the corresponding free energy 
of the system with $S$ kept constant. 
Bayes' theorem states that the joint probability
$P(\Gamma , S) ~=~ P(\Gamma |S) P(S) ~=~ P(S|\Gamma )P(\Gamma )$,
hence after some algebra, we obtain
\begin{equation}
P(S|struct)~=~ \sum_\Gamma P_{struct}(\Gamma ) P(S|\Gamma )
~=~ \sum_\Gamma P'_{struct}(\Gamma )
\exp ({-(H_S(\Gamma )-F_o(S))/T })
\label{eq:p'(s)}
\end{equation}
where
\begin{equation}
P'_{struct}(\Gamma ) = {P_{struct}(\Gamma ) \over
\sum_S \exp ({-(H_S(\Gamma)-F_o(S))/T})} .
\label{eq:p'}
\end{equation}
Because $P'_{struct}$ does not depend on the sequence $S$, we can
equally well regard it rather than $P_{struct}$ as given.
It can be thought
of as imposing an external clamping potential on the system, pushing it into the
correct structure,
$ P'_{struct}(\Gamma )  \equiv \exp ({-V_{ext}(\Gamma)/T})$ .
We will make an appropriate choice for the clamping potential, $V_{ext} (\Gamma)$
for each problem, based on physical reasoning rather than using 
(\ref{eq:p'}) directly.
Thus ({\ref{eq:p'(s)}) can be further reduced to
\begin{equation}
P(S|struct) \propto\exp ({-(F_{struct}(S)-F_o(S))/T})\equiv \exp ({-\Delta F/T})
\label{eq:minimize}
\end{equation}
where $F_{struct}(S)$ is the free energy {\it pushed}
into a certain structure by the clamping potential
\begin{equation}
\exp ({-F_{struct}(S)/T}) \equiv \sum_{\Gamma} \exp ({(-H_S(\Gamma)+V_{ext}(\Gamma))/T})
\end{equation}
The physical interpretation of this is clear. The
optimum sequence is the one that minimises the
difference $\Delta F$ between the unrestricted free
energy and the free energy clamped in the desired
structure. Intuitively this is reasonable because it picks out
a sequence that naturally wants to spend a lot of time in this
structure. This result is a generalisation of that
used in determining couplings in Boltzmann 
machines\cite{hertz}.

As a first example, we apply the formalism  above
to design copolymeric systems with desired lattice structures.
Leibler\cite{leibler} has developed a theory of microphase separation
for di-block copolymers.  Here we have further generalised this to $m$
blocks each block being made up entirely of A monomers or of B monomers,
the $i$th having a length $l_i,~ i=1\dots m$ . The fractional
length of the ith block is $f_i \equiv l_i/L$ where $L$ is the total
chain length.  The A and B monomers have an incompatibility, or Flory,
parameter $\chi$ giving the degree to which the A and B monomers wish
to segregate.  Leibler took as the order parameter $\Delta\rho$, the
ensemble average of the difference between the density of $A$'s 
and the average density of $A$'s. 
He was able to construct an expansion of the free energy in
terms of $\Delta\rho$ and calculated explicitly the expansion to fourth
order in terms of the underlying chemical structure of the chains, that
is, $l_1$, $l_2$, and $\chi$. This theory should work well near the
spinodal point for this system, because $\Delta\rho$ is small there.

To determine the stability of density variations with different crystallographic
symmetries, Leibler took $\Delta\rho$ to  be a periodic function of position
$\bf r$, 
\begin{equation}
 \Delta\rho ({\bf r}) = \sum_{j=1}^n \psi({\bf q_j}) \exp (i {\bf q}_j \cdot {\bf r}) ,
\label{eq:rho}
\end{equation}
choosing the ${\bf q}_j$'s, $j = 1,\dots, n$ to be the smallest 
non-zero reciprocal lattice
vectors of the lattice structure being considered. 
The magnitude, $q^*$, of the ${\bf q}_j$'s is taken to be the
wavevector at the spinodal point where divergent fluctuations first appear.
He further took the magnitude, but not the phase, of all the $\psi({\bf{
q}_j)}$'s to be equal.  By choosing different ${\bf q}_j$'s  and minimising
the resultant free energies he computed which crystal structure had the
lowest free energy. He was able to obtain a phase diagram for the
system as a function of $f_1$ and $\chi N$.  Besides the high
temperature disordered phase he found that the body centred cubic,
triangular (hex.), and lamellar (lam.) phases existed in different
regions of the phase diagram. Further work\cite{mayes1,olvera} extended
this treatment to triblocks.

It is convenient to take $V_{ext}(\Gamma)$ to be smoothly varying: 
\begin{equation}
 V_{ext} ({\bf r}) = -v \sum_{j=1}^n \exp (i {\bf q}_j \cdot {\bf r}) .
\end{equation}
The magnitude of $v$ adjusts the degree to which $\Delta\rho$
fluctuates.
A clamping potential with, for example, hexagonal symmetry is
the sum of three plane waves. Applying such a potential to the
unclamped free energy $F_0$, we obtain a clamped free energy
\begin{equation}
F(V_{ext}) = F_0 + \int \rho({\bf r})V_{ext}({\bf r}) d^3r ,
\end{equation}
which will tend to push the system into a phase with the 
symmetry of the external potential.

Unfortunately in the general formalism developed above
it is necessary that the clamping potential
$V_{ext}$ is very large when monomers stray from the desired
structure. This requirement leads to a large $\Delta\rho$ and hence
is incompatible with the limits of
validity of the free energy expansion of Leibler which is only
valid in the limit of weak segregation. Therefore we
need to consider complications that arise when only a weak clamping
potential is applied.

For small $V_{ext}$, the system will not always be pushed into the 
symmetry of the external potential.  In fact, for the values of $v$
that we use, the effect of adding $V_{ext}$ is only to slightly
enlarge the region of the phase diagram that has the symmetry of 
$V_{ext}$.  Our algorithm determines which phase the system is in 
by allowing the magnitudes of the $\psi({\bf q_j)}$'s to be 
unequal and then minimising with respect to them.  This allows 
the possibility of mixed phases with more than one type of 
symmetry.  For example if we consider the hexagonal phase
and write (\ref{eq:rho}) in terms of its 
wave vectors ${\bf q}_1$, ${\bf q}_2$, and ${\bf q}_3$,
\begin{equation}
\Delta\rho ({\bf r}) = \psi_{lam} \exp (i {\bf q}_1 \cdot {\bf r})
+\psi_{hex} (\exp (i{\bf q}_2 \cdot {\bf r})+
\exp (i{\bf q}_3 \cdot {\bf r})),
\label{eq:rhohex}
\end{equation}
then for $\psi_{lam} = \psi_{hex}$
we have a hexagonal structure ($n=3$ for the hex. phase), and for 
$\psi_{hex} = 0$ we have a lamellar structure ($n=1$ for the lam. 
phase).

For small $V_{ext}$, 
\begin{equation}
\Delta F \equiv F(V_{ext}) - F_0 ~=~ v (\partial\Delta F/\partial v)
= -v \sum_{j=1}^{n_c} \psi({\bf q_j}),
\label{eq:deltaf}
\end{equation}
where $n_c$ is the number of ${\bf q_j)}$'s that $V_{ext}$ has
in common with $\Delta\rho ({\bf r})$.  
From this it can be seen that $\Delta F$ does not do quite what one
would like. In general a clamping potential will have reciprocal
lattice vectors in common with more than one type of symmetry in
$\Delta\rho ({\bf r})$.  This means that $\Delta F$ will be 
lowered for other phases in addition to the one with the symmetry
of $V_{ext}$.  For example, for a clamping potential with
hexagonal symmetry $V_{hex}$, 
$\Delta F(V_{hex}) = F(V_{hex}) - F_0$ will be lowered for the
lamellar phase as well as the hexagonal. We would like a functional
which is more selective in order to design a hexagonal
material. The solution in this case is to subtract the lamellar
component as follows. Looking at (\ref{eq:rhohex}), we see that 
the appropriate functional to maximise is $\psi_{hex}$. We would
like to express  $\psi_{hex}$ in terms of free energy differences.
From (\ref{eq:deltaf}) 
$\psi_{hex} = -(\partial \Delta F(V_{hex})/\partial v)-
\partial \Delta F(V_{lam})/\partial v)/2 $. For small $v$
this is proportional to $F(V_{hex})-F(V_{lam})$. Therefore
minimising the difference in free energy between hexagonal
and lamellar clamping will bring the system into the hexagonal
phase. Fig. \ref{fig:hex}(a) shows $(F(V_{hex})-F(V_{lam}))$
as a function of the fraction $f$ for diblock copolymers at
$\chi ~=~ 20$. The minimum $f = 0.32$ occurs in the correct position
indicated by Leibler's theory. 

We also tested out this method for designing diblock bcc structures.
Fig. \ref{fig:hex}(b) shows $F(V_{bcc})-F(V_{hex})$
as a function of $f$ at $\chi ~=~ 20$. The minimum $f= .23$ 
also occurs in the correct position but there is also a secondary
small minimum at $f = .32$. This shows that
there is no guarantee that there will only be one minimum
for this minimisation function.

Now we turn to the problem of protein design. An interesting
approach has recently been proposed by
Shakhnovich and Gutin\cite{shakhnovich1,shakhnovich2}(SG), who proposed a method for solving this
problem for a simplified lattice model using a self avoiding chain. 
The model they employed has sequences
$\{ \sigma_i\}$ of
two possible monomer types that are given values $\pm 1$, for chains of length
$N$. This, plus the
positions of all the monomers $\{r_i\}$, completely describe the state
of the chain. We wish to find a sequence that causes the chain to fold
up into a desired structure, but we do not care about what the monomer
types end up being. The energy is
\begin{equation} 
E(\{\sigma_i\}, \{r_i\}) = 
{1\over 2} \sum_{i,j}^N (B_0 + B\sigma_i\sigma_j)\Delta({\bf r}_i-{\bf r}_j)
\label{eq:model}
\end{equation}
Here we will consider the case where $\Delta({\bf r}_i-{\bf r}_j) = 1$ if
${\bf r}_i$ and ${\bf r}_j$ are nearest neighbours, and is zero otherwise.
One sets $B < 0$, since this favours 
ferromagnetic ordering of the $\sigma$'s,
which means that the monomers will want to segregate, and $B_0 < 0$ 
since this provides 
an  attractive interaction between monomers causing the protein to collapse.  

Their method of sequence determination was to do a constrained
minimisation of the energy of the chain in sequence space. The
constraint was that the total magnetisation was held constant, in
practice, close to zero.  Note that without this constraint, all the
$\sigma$'s would become equal, and one would have a homopolymer which
does not have a well defined structure.  Unfortunately it appears that
even with constrained minimisation, sequences found do not necessarily
have to have the desired structures as the lowest energy
states\cite{whoops}. Even if the desired structure is a ground state,
there may be a large ground state degeneracy in which case the
structure is ill defined, as
in the case of the unconstrained minimisation just mentioned. 

The correct functional to minimise is $\Delta F$ defined in
(\ref{eq:minimize}).  If we specialise to the problem considered
here, the constraining potential is a delta function since our
structure is to be precisely determined. Calling the coordinates of the
desired structure $\{r^0_i\}$, the correct functional to minimise is,
according to (\ref{eq:minimize}),
\begin{equation} 
\Delta F ~=~ E(\{\sigma_i\}, \{r^0_i\}) - F_o(\{\sigma_i\}) .
\label{eq:correct}
\end{equation}
SG's method is a minimisation of only the first
term, with a constraint of constant total magnetisation. We will now 
argue that second term is not negligible and cannot be omitted. In
fact, we will see that a crude approximation gives an answer similar
to the constraint of constant magnetisation. But this approximation
is of dubious validity which is why it failed.

Take $B_0$ to be very large and negative so that all stable structures must be
globular with minimal surface area. The coordinates of our structure
$\{r^0_i\}$ are therefore constrained to be of this compact type 
and the term in the energy involving $B_0$ will not vary and can now be
ignored in the minimisation. The space of all conformations $\{r_i\}$
we need to consider are also compact conformations of the same overall
shape, but different internal arrangements. We now
expand out $F(\{\sigma_i\})$ keeping only the lowest order cumulant.
$F(\{\sigma_i\}) \approx \langle E\rangle +~ constant$. 
The angled brackets denote an average that is equally weighted 
over all compact conformations with minimal surface area. 
Ignoring constant terms, this gives 
$ F_o(\{\sigma_i\}) \approx {B\over 2} \sum_{i,j}^N  
\sigma_i\sigma_j\langle\Delta({\bf r}_i-{\bf r}_j)\rangle $ so
\begin{equation}
\Delta F \approx {B\over 2} \sum_{i,j}^N  
[ \Delta({\bf r}_i-{\bf r}_j)-\langle\Delta({\bf r}_i-{\bf r}_j)
\rangle ] \sigma_i\sigma_j
\label{eq:approx}
\end{equation} 
One can easily show that the nearest neighbour interactions  along the
backbone of the chain cancel, 
because the probability
that monomers $i$ and $i+1$ are next to each other is unity. This
must happen because this kind of interaction is the same for all
configurations and consequently cannot play a role in choosing
the optimum $\sigma$'s.
We now find an approximate functional form for
$\langle\Delta({\bf r}_i-{\bf r}_j)\rangle$. Since the chain is compact
there is a short screening length. We
therefore expect random walk correlations when $|i-j|$ is greater than
a few lattice spacings. However when $|i-j|^{1/2}$ becomes of order the
diameter of the protein, the conformations cease to look like random
walks as the protein is compact. This crossover 
corresponds to $|i-j| \sim N^{2/3}$. For scales larger than this the
correlation function should be almost constant. Therefore
\begin{equation}
\langle\Delta({\bf r}_i-{\bf r}_j)\rangle \sim
\left\{
\begin{array}{l}
|i-j|^{-3/2}\;\;\mbox{for}\;1 << |i-j| << N^{2/3}\; , \\
1/N\;\;\mbox{for}\; |i-j| >> N^{2/3} .
\end{array}\right.
\end{equation}
Therefore in this approximation, $F(\{\sigma_i\})$ looks like a one
dimensional Ising model with the above long range interaction. 

If we ignore the variation of $\langle\Delta({\bf r}_i-{\bf r}_j)\rangle$ with
$|i-j|$, i.e. $\langle\Delta({\bf r}_i-{\bf r}_j) \rangle = 1/N$  so
$\Delta F_{MF} =  E(\{\sigma_i\}, \{r^0_i\}) - B(\sum_i^N \sigma_i)^2/(2N)$,  
then $F_o$ 
gives an infinite range mean field contribution that is
antiferromagnetic, and hence acts as a ``soft'' constraint 
favouring a total magnetisation of zero. To have a total magnetisation
of zero, one must introduce a domain wall, which increases 
$E(\{\sigma_i\}, \{r^0_i\})$ by of order $N^{2/3}$, but this is
more than compensated by the gain in free energy of the second term
which is of order $N$. Hence in this limit we recover the approximation 
of SG. However this is a rather drastic approximation.
Even within the first order expansion derived above, one is {\em not}
justified in neglecting the shorter distance variation of 
$\langle\Delta({\bf r}_i-{\bf r}_j)\rangle$ because, 
within this mean field approximation,
there is a spurious degeneracy in the $\sigma$'s that minimise
$\Delta F_{MF}$.
This is because
the three dimensional arrangement of the $\sigma_i$'s namely $\sigma({\bf r})$
is {\em independent} of the desired conformation $\{r^0_i\}$. That is,
to find the correct sequence, it is not necessary to consider the
different internal arrangements of the chain inside the compact cluster,
as they all give an identical $\Delta F_{MF}$. After the 
minimisation of $\Delta F_{MF}$
has been performed once, the three dimensional arrangement of the $\sigma$'s
do not change when the desired conformation is changed.

However a non-constant
$\langle\Delta({\bf r}_i-{\bf r}_j)\rangle$ breaks this degeneracy. It adds
antiferromagnetic couplings along the backbone of the chain. This
means that the domain wall should tend to orient itself roughly 
perpendicular to the direction of the backbone of the chain to 
satisfy the antiferromagnetic couplings. The contribution to $\Delta F$
due to the $|i-j|^{-3/2}$ decay of $\langle\Delta({\bf r}_i-{\bf r}_j)\rangle$
is substantial and cannot  be neglected. We can easily estimate it
for the case of a desired structure $\{r^0_i\}$ that is in a 
typical random configuration. For a piece of  arclength $s \equiv 
|i-j| = N^{2/3}$ the contribution to $\Delta F$ is of order 
$B s^2/s^{3/2} = B L^{1/3}$. But there are $L/L^{2/3}$ such pieces, giving
a total contribution of order $B L^{2/3}$. This precisely the same
order as the energy of the domain wall and therefore must be considered
in doing protein design.

In conclusion, we have developed a method to design molecules that
will self assemble into a desired structure.
We used this to design block copolymers
with desired structural characteristics, within 
the framework of Leibler's mean field theory.

We also note that the method for design described
above should also work for the problem of protein design,
for which no trustworthy methods have been devised so far.
It would be interesting to see how well the approximate minimisation
function, given by (\ref{eq:approx}),
designs stable proteins. 

One of us (T.K.) wishes to thank Mark Dinan, Hemant Bokil
and Doug Williams for useful discussions. We would also
like to thank A.P. Young for useful comments and a 
critical reading of the manuscript.
This work is supported by NSF grant number DMR-9419362
and acknowledgement
is made to the Donors of the Petroleum Research Fund, administered
by the American Chemical Society for partial support of this research.

\newpage
\begin{figure}[tbh]
\begin{center}
\
\psfig{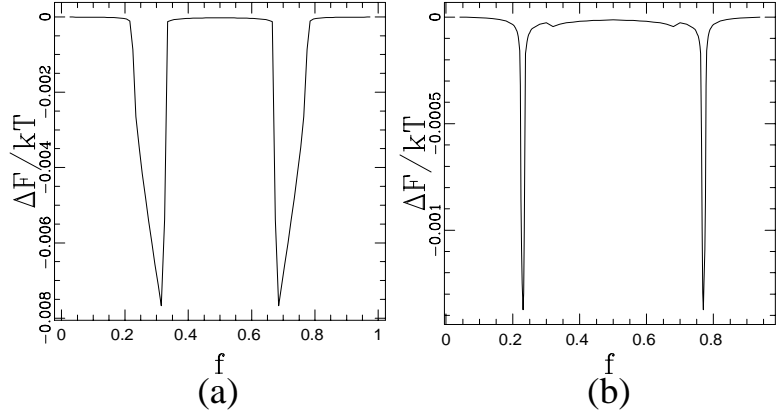}
\end{center}
\caption[Hexagonal Minimum] {(a) $F(V_{hex})-F(V_{lam})$
and (b)$F(V_{bcc})-F(V_{hex})$, 
as a function of the fraction $f$ for diblock copolymer
design. The minima give the best choice for
the design of hexagonal material and bcc material, respectively.}
\label{fig:hex}
\end{figure}

\end{document}